\documentclass[aps,prl,preprint]{revtex4}

\usepackage{epsfig}

\draft

\begin{document}
\author{Hong-shi Zong$^{1,2,3}$, Shi Qi$^{1}$, Wei Chen$^{1}$, Wei-min Sun$^{1}$, and En-guang Zhao$^{2,3}$}
\address {$^{1}$ Department of Physics, Nanjing University, Nanjing 210093, P. R. China}
\address{$^{2}$ CCAST(World Laboratory), P.O. Box 8730, Beijing 100080, P. R. China}
\address{$^{3}$ Institute of Theoretical Physics, Academia Sinica, P.O. Box 2735, Beijing 100080, P. R. China}

\title{Pion susceptibility of the QCD vacuum from an effective quark-quark interaction}

\begin{abstract}
A modified method for calculating the pion vacuum susceptibility from an effective quark-quark interaction model
is derived. Within this approach it is shown that the pion vacuum susceptibility is free of ultraviolet divergence and is much smaller than the previous estimations.

\bigskip


\bigskip

E-mail: zonghs@chenwang.nju.edu.cn.

\bigskip


\end{abstract}

\maketitle
The vacuum condensate is known to play the essential role for characterizing the non-perturbative aspect of the QCD 
vacuum. In fact, the gauge invariant vacuum condensates such as the chiral quark condensate, the two gluon condensate and the mixed quark gluon condensate are well known examples studied extensively so far[1,2].  Meanwhile susceptibilities of vacuum are also important quantities of strong interaction physics. They directly enter in the determination of hadron properties in the QCD sum rule external field approach[3--5]. In particular, the strong and parity--violating pion--nucleon coupling depends crucially upon $\chi^{\pi}$, the $\pi$ susceptibility[6]. The previous estimation for the value of pion susceptibility were obtained by a QCD sum rule three-point formalism[7]. It is the aim of this letter to consider $\pi$ vacuum susceptibility in the framework of the global color symmetry model(GCM). Though GCM[8] violate local color $SU(3)_{C}$ gauge invariance and renormalizability, they provide a successful description of various nonperturbative aspects of strong interaction physics and hadronic phenomena at low energies[8-22]. A few more recent review articles on this topic can be found in Refs.[23,24].

The GCM generating functional for massless quarks in Euclidean space
can be written as[8]
\begin{equation}
Z[\bar{\eta}, \eta] = \int {\cal{D}} B^{\theta}(x,y) \exp(-S [ \bar{\eta},
\eta, B^{\theta}(x,y)] ) \, ,
\end{equation}
with
\begin{eqnarray}
S[\bar{\eta}, \eta, B^{\theta}(x,y)] & = & - \mbox{Tr} \ln \left[\rlap/
\partial \delta (x-y) + \frac{1}{2} \Lambda^{\theta} B^{\theta}(x,y) \right]
\nonumber \\
& & + \int\int d^4 x d^4 y\left[\frac{B^{\theta}(x,y)
B^{\theta}(y,x)}{2 g^2 D(x-y)}
+ \bar{\eta}(x) {\cal{G}}(x,y;[B^{\theta}])\eta(y) \right],
\end{eqnarray}
where $B^{\theta}(x,y)$ is an auxiliary bilocal field introduced as in Ref.[8] and the matrices $\Lambda^{\theta} = D^{a} F^{b} C^{c}$ is determined by Fierz
transformation in Dirac, flavor and color space(more detail can be found in Ref.[8]); $D^{ab}_{\mu\nu}(x-y)$ denotes the effective gluon two-point green function. Because the form of $D^{ab}_{\mu\nu}(x-y)$ in the infrared region is unknown, one often treat the $D^{ab}_{\mu\nu}(x-y)$ as the GCM input parameter, which, as we will discuss later, is chosen to reproduce certain aspects of low energy hadronic physics.  For convenience we will use the Feynman like gauge $ D_{\mu\nu}^{ab} (x-y)=\delta_{\mu\nu}\delta^{ab} D(x-y)$ from now on.

The saddle-point of the action is defined as
$\left.\delta S[\bar{\eta}, \eta, B^{\theta}(x,y)]/\delta B^{\theta}(x,y)
\right\vert _{\eta =\bar{\eta}
= 0} =0$ and is given by
\begin{equation}
B^{\theta}_{0}(x-y)=g^2 D(x-y) tr[\Lambda^{\theta} {\cal{G}}_{0}(x-y)],
\end{equation}
where ${\cal{G}}_{0}$ stands for ${\cal{G}}[B^{\theta}_{0}]$ and the 
trace in Eq.(3) is to be taken in Dirac and color space, whereas the flavor trace has been separated out.

We will calculate the vacuum condensates from the saddle-point expansion, that is, we will work at the mean field level. This is consistent with the large $N_{C}$ limit in the quark fields for a given model gluon two-point function. In the mean field approximation, the field $B^{\theta}(x-y)$ is substituted by their vacuum $B^{\theta}_{0}(x-y)$. Under this approximation, the dressed quark propagator $G(x-y)\equiv {\cal{G}}_{0}(x-y)$ in GCM has the decomposition
\begin{equation}
G^{-1}(p)\equiv i\gamma\cdot p+\Pi(p)\equiv i\gamma\cdot p A(p^2)+B(p^2)
\end{equation}
with the self-energy dressing of the quarks $\Pi(p)$ through the definition:
\begin{equation}
\Pi(p)\equiv \frac{1}{2} \Lambda^{\theta}B^{\theta}_{0}(p)=\int d^4x e^{ip\cdot x}\left[\frac{1}{2} \Lambda^{\theta}B^{\theta}_{0}(x)\right] =i\gamma\cdot
p[A(p^2)-1]+B(p^2) ,
\end{equation}
where the self energy functions $A(p^2)$ and $B(p^2)$ are
determined by the rainbow Dyson-Schwinger equation[8]
\begin{equation}
[A(p^2)-1]p^2=\frac{8}{3}\int \frac{d^{4}q}{(2\pi)^4}g^2 D(p-q)
\frac{A(q^2)p\cdot q}{q^2A^2(q^2)+B^2(q^2)},
\end{equation}
\begin{equation}
B(p^2)=\frac{16}{3}\int \frac{d^{4}q}{(2\pi)^4}g^2 D(p-q)
\frac{B(q^2)}{q^2A^2(q^2)+B^2(q^2)}.
\end{equation}
In order to get the numerical solution of $A(p^2)$ and $B(p^2)$, one often use model forms for gluon two-point function as input in Eqs.(6,7). As a typical example, we choose $g^2D(q^2)=4\pi^2d\frac{\lambda^2}{q^4+\Delta}$ with $d=\frac{12}{27}$, and three sets of different parameters for $\lambda$ and $\Delta$, which are adjusted to reproduce the pion decay constant in the chiral limit $f_{\pi}$=$87 MeV$ and the chiral low energy coefficients[14,25].

Here we want to stress that the $B(p^2)$ in Eqs.(6,7) has two qualitatively distinct solutions. The ``Nambu-Goldstone'' solution, for which
\begin{equation}
B(p^2)\neq 0,
\end{equation}
describes a phase in which: 1) chiral symmetry is dynamically broken. Because one has a nonzero quark mass function; and 2) the dressed quarks are confined, because the propagator described by these functions does not have a Lehmann representation[9]. In ``Nambu-Goldstone'' phase, the vacuum configuration $ B^{\theta}_{0}(x-y)$ in GCM(at the mean field approximation) can be regarded as a good approximation to the ``exact'' vacuum in QCD. The alternative ``Wigner'' solution, for which
\begin{equation}
B(p^2)\equiv 0,
\end{equation}
describes a phase in which chiral symmetry is not broken and the dressed-quarks are not confined. 

With these two ``phase'' characterized by qualitatively different quark propagator, the GCM can be used to calculate the vacuum condensates and
susceptibilities as we will show below.

In the external field of QCD sum rule two--point method, one often encounters
the quark propagator in the presence of the
$J^{\Gamma}(y)=\bar{q}(y)\Gamma q(y)$ current($\Gamma$ stands for the
 appropriate combination of Dirac, flavor and color matrices).
\begin{eqnarray}
S^{cc'\Gamma}_{\alpha\beta}(x)=\langle\tilde{0}|T[q^{c}_{\alpha}(x)\bar{q}^{c'}_{\beta}(o)]\tilde{0}
\rangle_{J^{\Gamma}}=S^{cc'\Gamma,PT}_{\alpha\beta}(x)
+S^{cc'\Gamma,NP}_{\alpha\beta}(x),
\end{eqnarray}
where $S^{\Gamma,PT}_{q}(x)$ is the quark propagator coupled perturbatively
to the current and $S^{\Gamma,NP}_{q}(x)$ is the nonperturbative quark
propagator in the presence of the external current $J^{\Gamma}$(one should
note that the external current should be taken to be $J^{\Gamma}\phi_{\Gamma}$, where $\phi_{\Gamma}$
is the value of the external field. In what follows, to simplify the notation, we will take the $\phi_{\Gamma}$=1, which does not affect our results).
The vacuum susceptibility $\chi^{\Gamma}$ in the QCD sum rule two--point  external field treatment can be defined as[7]
\begin{eqnarray}
S^{cc'\Gamma,NP}_{\alpha\beta}(x)=\langle\tilde{0}|:q^{c}_{\alpha}(x)
\bar{q}^{c'}_{\beta}(0):|\tilde{0}\rangle_{J^{\Gamma}}
=-\frac{1}{12}\Gamma_{\alpha\beta}\delta_{cc'}
\chi^{\Gamma}H(x)\langle 0|:\bar{q}(0)q(0):|0\rangle ,
\end{eqnarray}
where the phenomenological function $H(x)$ represents the nonlocality of the
two quark nonlocal condensate. Note that H(0)=1.

The presence of external field implies that $S^{cc'\Gamma}_{\alpha\beta}(x)$
is evaluated with an additional term $\Delta L\equiv -J^{\Gamma}\cdot\phi_{\Gamma}$ added to the usual QCD Lagrangian. In three point method of QCD sum rule, if one takes only a linear external
field approximation, the $S^{cc'\Gamma,NP}_{\alpha\beta}(x)$ in Euclidean space
is given by[7]
\begin{equation}
S^{cc'\Gamma,NP}_{\alpha\beta}(x)=
\int d^{4}y \langle\tilde{0}|:q^{c}_{\alpha}(x)\bar{q}^{e}(y)
\Gamma q^{e}(y)\bar{q}^{c'}_{\beta}(0):|\tilde{0}\rangle .
\end{equation}

Using Eq.(11), (12) and note that $H(0)=1$ we have
\begin{equation}
-\frac{1}{12}\delta_{cc'}\Gamma_{\alpha\beta}\chi^{\Gamma}
\langle \tilde{0}|:\bar{q}(0) q(0):|\tilde{0} \rangle
=\int d^{4}y \langle \tilde{0}|:q^{c}_{\alpha}(0)\bar{q}^{e}(y)
\Gamma q^{e}(y)\bar{q}^{c'}_{\beta}(0):|\tilde{0}\rangle .
\end{equation}

Multiplying Eq.(13) by $\Gamma_{\beta\alpha}\delta_{cc'}$, we get
\begin{equation}
\chi^{\Gamma}a
=-\frac{16\pi^2}{tr_{\gamma}(\Gamma\Gamma)}\int d^{4}y
\langle \tilde{0}|:\bar{q}^{c}(0)\Gamma q^{c}(0)
\bar{q}^{e}(y)\Gamma q^{e}(y):|\tilde{0}\rangle ,
\end{equation}
with $a$=-$(2\pi)^2\langle 0|:\bar{q}(0) q(0):|0 \rangle$. Eq.(14) shows that the vacuum susceptibilities originates from the nonlocal four quark condensate[7]. Therefore, in order to calculate the vacuum susceptibilities, one must know how to define and calculate the vacuum condensates in advance.

Let us first recall the definition of vacuum condensate in QCD sum rule. In QCD sum rule, one often postulate that quark propagators are modified by the long-range confinement part of the QCD; but the modification is soft in this sense that at short distance the difference between exact and perturbative propagators vanishes.

To formalize this statement, one can write the ``exact'' propagator $G(x)$ as a vacuum expectation of a T-product of fields in the ``exact'' vacuum $|\tilde{0}\rangle$
\begin{equation}
G_{ij}(x,y)\equiv\langle\tilde{0}|T[q_i(x)\bar{q}_j(y)]|\tilde{0}\rangle.
\end{equation}

According to the Wick theorem, one can write the T-product as the sum
\begin{equation}
T[q_i(x)\bar{q}_j(y)]\equiv\underbrace{q_i(x)\bar{q}_j(y)}+:q_i(x)\bar{q}_j(y):
\end{equation}
of the ``pairing'' and the ``normal'' product. The ``pairing'' is just the expectation value of the T-product over the perturbative vacuum $|0\rangle$
\begin{equation}
\underbrace{q_i(x)\bar{q}_j(y)}\equiv\langle 0|T[q_i(x)\bar{q}_j(y)]|0\rangle\equiv G_{ij}^{pert}(x,y).
\end{equation}
i.e., the perturbative propagator. By this definition, we have the following two quark vacuum condensate $\langle\tilde{0}|:\bar{q}(x)q(y):|\tilde{0}\rangle$:
\begin{eqnarray}
&&\langle \tilde{0}|:\bar{q}_i(x)q_j(y):|\tilde{0}\rangle\equiv
\langle\tilde{0}|T[\bar{q}_i(x)q_j(y)]|\tilde{0}\rangle
-\langle 0|T[\bar{q}_i(x)q_j(y)]|0\rangle \\
&&\equiv (-)\left[G_{ji}(y,x)-G_{ji}^{pert}(y,x)\right]\equiv (-)\Sigma_{ji}(y,x)
=(-)\int\frac{d^4 q}{(2\pi)^4}e^{i q\cdot (y-x)}\left[G(q^2)-G^{pert}(q^2)\right]_{ji},
\nonumber
\end{eqnarray}
where $\Sigma(x,y)\equiv G(x,y)-G^{pert}(x,y)$. 

Similarly, we have the four quark vacuum condensate:
\begin{eqnarray}
&& \langle \tilde{0}|:\bar{q}(x)\Lambda^{(1)}q(x)\bar{q}(y)\Lambda^{(2)}q(y):|\tilde{0}\rangle \nonumber\\
&&=\langle \tilde{0}|T\left[\bar{q}(x)\Lambda^{(1)}q(x)
\bar{q}(y)\Lambda^{(2)}q(y)\right]|\tilde{0}\rangle
-\langle 0|T\left[\bar{q}(x)\Lambda^{(1)}q(x)
\bar{q}(y)\Lambda^{(2)}q(y)\right]|0\rangle \nonumber\\
&&-\langle \tilde{0}|:\bar{q}(x)\Lambda^{(1)}\underbrace{q(x)
\bar{q}(y)}\Lambda^{(2)}q(y):|\tilde{0}\rangle
-\langle \tilde{0}|:\underbrace{\bar{q}(x)\Lambda^{(1)}q(x)
\bar{q}(y)\Lambda^{(2)}q(y)}:|\tilde{0}\rangle\nonumber\\
&&-\langle \tilde{0}|:\underbrace{\bar{q}(x)\Lambda^{(1)}q(x)}
\bar{q}(y)\Lambda^{(2)}q(y):|\tilde{0}\rangle
-\langle \tilde{0}|:\bar{q}(x)\Lambda^{(1)}q(x)
\underbrace{\bar{q}(y)\Lambda^{(2)}q(y)}:|\tilde{0}\rangle\nonumber\\
&&=-\left\{tr[\Sigma(y,x)\Lambda^{(1)}
\Sigma(x,y)\Lambda^{(2)}]
-tr[\Sigma(x,x)\Lambda^{(1)}]
tr[\Sigma(y,y)\Lambda^{(2)}]\right\},
\end{eqnarray}
here the $\Lambda^{(i)}$ stands for an operator in Dirac and color space. 

Based on the above statement, in order to calculate quark vacuum condensates, one must know not only the ``exact'' but also the ``perturbative'' quark propagator in advance. The calculation of ``exact'' quark propagator in GCM(at the mean field approximation) has been given by Eqs.(6,7) with Eq.(8), the key question left now is how to treat consistently the ``perturbative'' quark propagator in GCM. Because the form of perturbative quark propagator in the low momentum region is unknown(one only know the large momentum behavior of perturbative quark propagator), it is practically very difficulty to provide a $G^{pert}$ for whole momentum region in QCD. However, in phenomenological application, one may proceed by making some assumptions regarding the ``perturbative'' quark propagator.

As a first ``guess'' for a perturbative quark propagator, one would simply take the bare quark propagator($A(p^2)$=1, $B(p^2)$=0) or the perturbative quark propagator at one loop approximation. However, numerical calculation for the mixed quark gluon condensate in GCM shows that one can not get a reasonable numerical value for the mixed quark gluon condensate from this simple ``guess''. In order to obtain the reasonable result for the mixed quark gluon condensate, the authors in Ref.[26] identified the ``perturbative'' vacuum with the ``Wigner'' vacuum(no chiral symmetry breaking and the dressed-quarks are not confined), in contrast to the ``Nambu-Goldstone'' vacuum corresponding to dynamical symmetry breaking and the dressed-quarks are confined.  

In ''Wigner'' phase(the $B'(p^2)\equiv 0$), the Dyson-Schwinger equation(6,7) reduces to:
\begin{equation}
[A'(p^2)-1]p^2=\frac{8}{3}\int \frac{d^{4}q}{(2\pi)^4}g^2 D(p-q)
\frac{p\cdot q}{q^2A'(q^2)},
\end{equation}
where the solution $A'(p^2)$ in Eq.(20) denotes the quark propagator propagating in the ``Wigner'' vacuum(It also sums infinitely many effective one-gluon exchange diagrams, as does the solution in ``Nambu-Goldstone'' phase).  If one use the vacuum configuration $B'^{\theta}_{0}$ of the ``Wigner'' mode(at the mean field approximation) to model the ``perturbative'' vacuum in QCD(As it did in Ref.[26]), the ``perturbative'' quark propagator in GCM can be written as $G^{pert}(q^2)=\frac{-i\gamma\cdot q}{A'(q^2)q^2}=-i\gamma\cdot q C(q^2)$. It should be noted that the ``perturbative'' propagator here is only a notation(In order to be consisted with the one used in QCD sum rule) which denotes the dressed quark propagator in ``Wigner'' phase.  Numerical studies also show that for $q^2\gg \Lambda^2_{QCD}$, one has
\begin{equation}
G(q^2)-G^{pert}(q^2)=\frac{-i\gamma\cdot q A(q^2)+B(q^2)}{A^2(q^2)q^2+B^2(q^2)}+i\gamma\cdot q C(q^2)\to 0,
\end{equation}
i.e., at short distance the difference between exact and perturbative propagators tends to zero(see Eq.(27) below). This result is just what one expected in advance. It can be seen from the following figures (fig.1--fig.2).

\epsfxsize=3.0in \epsfbox{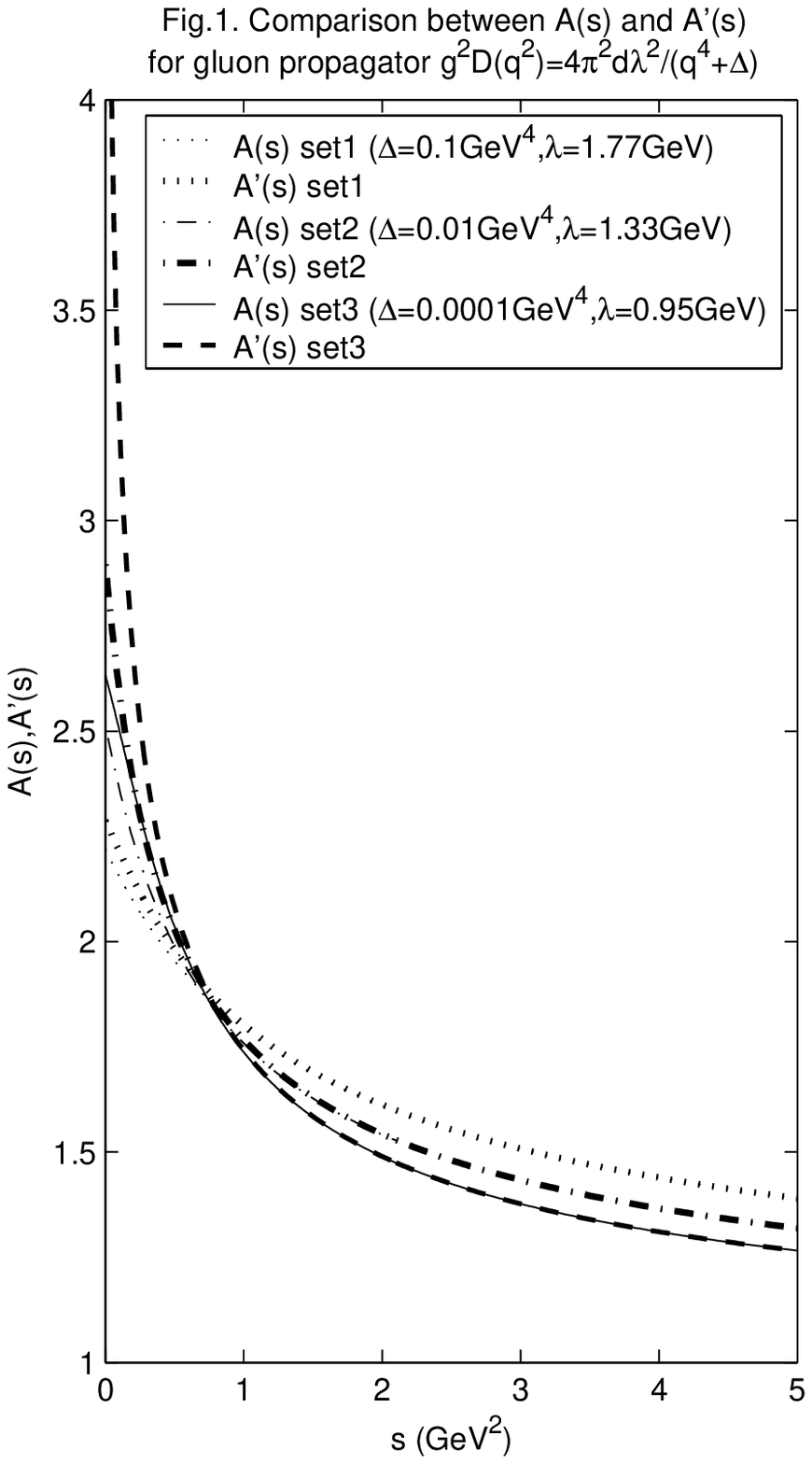}
\epsfxsize=3.0in \epsfbox{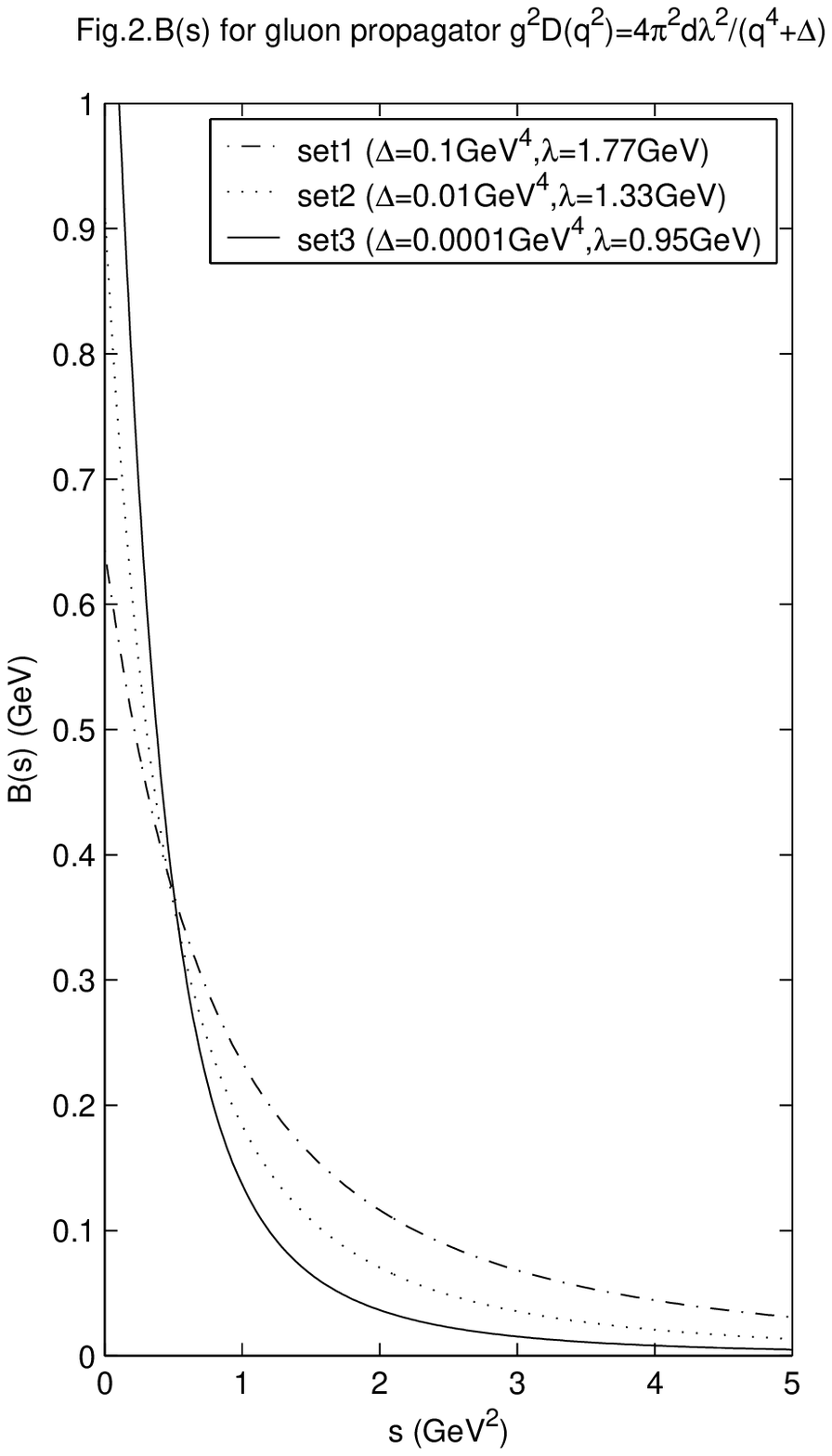}

Once the ``exact'' and ``perturbative'' quark propagator in our model are determined,
one can calculate the $\langle\tilde{0}|:\bar{q}q:|\tilde{0}\rangle$,
$\langle\tilde{0}|:\bar{q} \Lambda^{(1)} q\bar{q} \Lambda^{(2)} q:|\tilde{0}\rangle$ vacuum condensate at the mean field level. In particular we obtain the two quark condensate $\langle\tilde{0}| :\bar{q}q:|\tilde{0}\rangle$ in the chiral limit;
\begin{eqnarray}
\langle\tilde{0}| :\bar{q}q:|\tilde{0}\rangle
=(-)tr[\Sigma (x,0)]|_{x=0}
=(-)\lim_{\mu\to\infty}\frac{3}{4\pi^2}\int^{\mu}_{0} ds s
\frac{B(s)}{sA^2(s)+B^2(s)}.
\end{eqnarray}
It should be noted that Eq.(22) is appropriate only when we use an effective gluon propagator; i.e., $g^2D(q^2)$, falls off like $1/q^4$ or more quickly(for example using the form of $g^2D(q^2)$=$4\pi^2d\frac{\lambda^2}{q^4+\Delta}$). In this case the model does not need renormalization and the chiral quark condensate defined in Eq.(22) is free of UV divergence(There are many such models, the instanton liquid model among them). However, if we take into account the asymptotic behavior of the gluon propagator(for example using the form of Eqs.(25,26) below), the renormalization is necessary, and the naive formula Eq.(22) for the condensate has to be multiplied by a renormalization constant(more detail can be found in Ref.(12) and (16)).  

Returning to the calculation of vacuum susceptibilities, in the case of tensor current($\Gamma=\sigma_{\mu\nu}$), from Eqs.(14), (19) and the fact $tr_{\gamma}(\sigma_{\mu\nu}\sigma_{\mu\nu})=48$, we obtain the tensor vacuum susceptibility $\chi^{T}a$:
\begin{equation}
\chi^{T}a=(48\pi^2)
\int \frac{d^4p}{(2\pi)^4}\left[\frac{B(p^2)}{A^2(p^2)p^2+B^2(p^2)}\right]^2
=3\lim_{\mu\to\infty}\int^{\mu}_{0} sds\left[\frac{B(s)}{A^2(s)s+B^2(s)}\right]^2.
\end{equation}
The calculation of the tensor vacuum susceptibility in the framework of GCM has been given in Ref.[27].

Similarly, in the case of pseudoscalar current, we have the pion susceptibility 
$\chi^{\pi}a$:
\begin{eqnarray}
\chi^{\pi}a
&&=\frac{3}{\pi^2}\int d^4 p\left\{\left[
\frac{A(p^2)}{A^2(p^2)p^2+B^2(p^2)}-C(p^2)\right]^2 p^2+\left[\frac{B(p^2)}{A^2(p^2)p^2+B^2(p^2)}\right]^2\right\} \nonumber\\
&&=3\lim_{\mu\to\infty}\int^{\mu}_{0} sds\left\{\left[\frac{A(s)}{A^2(s)s+B^2(s)}-C(s)\right]^2s+\left[
\frac{B(s)}{A^2(s)s+B^2(s)}\right]^2\right\},
\end{eqnarray}
which is UV convergent. It is reasonable due to the vacuum condensates reflecting the IR behavior of QCD. However, if we do not adopt an adequate subtraction mechanism(i.e., $C(p^2)$=0) to calculate the $\pi$ vacuum susceptibility, Eq.(24) would be UV divergent. In addition, it should be noted that the calculation of nonlocal four quark condensate $\langle \tilde{0}|:\bar{q}^{c}(0)\Gamma q^{c}(0) \bar{q}^{e}(y)\Gamma q^{e}(y):|\tilde{0}\rangle$ here is quite different from the previous estimation[7]. In Ref.[7], the authors did not adopt subtraction mechanism to calculate nonlocal four-quark condensate, instead they used vacuum saturation hypothesis and retained only the contribution of the scalar condensates(more details can be found in Ref.[7]).  

From Eq.(22) and (24), one can calculate the two-quark condensate and the pion vacuum susceptibility. In order to show the importance of the proper choice of the perturbative quark propagator, we use separately the ``Wigner'' solution of Eq.(20) and the bare quark propagator for the perturbative propagator to calculate the pion vacuum susceptibility $\chi^{\pi}a$ and $\chi'^{\pi}a$ in Eq.(24). In Table.I, the results for the two quark vacuum condensate and $\pi$ vacuum susceptibilities for three different parameters sets of the model gluon two-point function are displayed.

\begin{table}[ht]
\caption{\label{Tablechi2}
The two quark condensate and $\pi$ vacuum susceptibility for $g^2D(q^2)$=$4\pi^2d\frac{\lambda^2}{q^4+\Delta}$}
\begin{center}
\begin{tabular}{ccccc}\hline\hline
$\Delta[GeV^4]$~~~~~~~ & $\lambda[GeV]$ ~~~~~~~ &
 $-(\bar{q}q)^{\frac{1}{3}}$~$(MeV)$~~~~~~~~ & $\chi^{\pi}a$~$(GeV^2)$ ~~~~~~~~ & $\chi'^{\pi}a$~$(GeV^2)$\\ \hline
$10^{-1}$ ~~~~~~~ & 1.77 ~~~~~~~~    &   291  ~~~~~~~~~           &  0.109 ~~~~~~~~~           & 6.402\\
$10^{-2}$ ~~~~~~~ & 1.33 ~~~~~~~~    &   250  ~~~~~~~~            &  0.089~~~~~~~~~            &4.603\\
$10^{-4}$ ~~~~~~~ & 0.95 ~~~~~~~     &   217  ~~~~~~~~            &  0.071~~~~~~~~~           &3.583\\ \hline\hline
\end{tabular}
\end{center}
\end{table}
Table.I show that our two quark condensate is compatible with the ``standard'' value of $-(250 MeV)^3$ in QCD sum rule, whereas the $\pi$ vacuum susceptibility $\chi^{\pi}a$ is much smaller than the value $\chi^{\pi}a\simeq (1.7-3.0)~GeV^2$ obtained within a phenomenological approach[7]. In addition, the huge difference between the $\chi^{\pi}a$ and $\chi'^{\pi}a$ clearly shows the importance of the proper choice of the perturbative quark propagator. Due to the subtraction of bare quark propagator(or the perturbative quark propagator at one loop) is not appropriate for the calculation of the mixed quark gluon condensate[26],  we take the $\chi^{\pi}a$ as our prediction for the $\pi$ vacuum susceptibility.

As is shown in Ref.[26], in a correct treatment of vacuum condensate, the cutoff $\mu$ should be taken to be infinity. Because the calculation is numerical, so one have to use a very large but finite value of upper limit of integration. In our numerical calculation, the upper limit is chosen to be $500~GeV^2$, which is large enough to ensure that the calculated value is independent of the choice of the upper limit of integration. This result can be seen from Table II.

To check if the small value of the pion susceptibility is due to a special form for the effective gluon propagator, other two popular forms of the gluon propagator[14,25] have been employed. They are model 1:
\begin{equation}
g^2D^{(1)}(q^2)=g^2D^{(1)}_{IR}(q^2)+g^2D_{UV}(q^2)\equiv 3\pi^{2}\frac{\lambda^2}{\Delta^2}e^{-\frac{q^2}{\Delta}}+\frac{4\pi^2 d}{q^2ln\left(\frac{q^2}{\Lambda^2_{QCD}}+e\right)},
\end{equation}
and model 2:
\begin{equation}
g^2D^{(2)}(q^2)=g^2D^{(2)}_{IR}(q^2)+g^2D_{UV}(q^2)\equiv 4\pi^2 d\frac{\lambda^2}{q^4+\Delta}+\frac{4\pi^2 d}{q^2ln\left(\frac{q^2}{\Lambda^2_{QCD}}+e\right)}.
\end{equation}

\begin{table}[ht]
\caption{\label{Tablechi2}
The sensitivity of the vacuum condensates to the choice
of the upper limit of integration $\mu$ in Eqs.(22) and (24)
for $g^2D(q^2)$=$4\pi^2d\frac{\lambda^2}{q^4+\Delta}$($\Delta=10^{-2}GeV^4$,~~$\lambda=1.33GeV$)}
\begin{center}
\begin{tabular}{ccc}
\hline\hline
$\mu$~($GeV^2$)~~~~~~~~~~~ & (--)$\langle\tilde{0}| :\bar{q}q:|\tilde{0}\rangle^\frac{1}{3}~(MeV)$~~~~~~~~~~
& $\chi^{\pi}a$~$(GeV^2)$ \\ \hline
1  ~~~~~~~~~~~      &    179   ~~~~~~~~~~~              & 0.083  \\
10 ~~~~~~~~~~~      &    241   ~~~~~~~~~~~             &  0.089\\
100~~~~~~~~~~~      &    249   ~~~~~~~~~~~              & 0.089\\
300~~~~~~~~~~~      &    250   ~~~~~~~~~~~             &  0.089\\
500~~~~~~~~~~~      &    250   ~~~~~~~~~~~              & 0.089\\\hline\hline
\end{tabular}
\end{center}
\end{table}
The term $D_{IR}(q^2)$, which dominates for small $q^2$, simulates the infrared enhancement and confinement. The other term $D_{UV}(q^2)$, which dominates for large $q^2$, is an asymptotic ultraviolet(UV) tail match the known one-loop renormalization group result with $d=[12/(33-2N_{f})]=12/27$, $\Lambda_{QCD}=200~MeV$. The model parameters $\lambda$ and $\Delta$ are adjusted to reproduce chiral low energy coefficients and the pion decay constant in the chiral limit as previously did. The calculated results are shown in Table III. 

\begin{table}[ht]
\caption{\label{Tablechi2}
$\chi^{\pi}a$ for Models 1 and 2 with three sets of different parameters, respectively.} 
\begin{center}
\begin{tabular}{ccc|ccc}\hline\hline 
\multicolumn{3}{c|}{Model 1}&\multicolumn{3}{c}{Model 2}\\ \hline
$\Delta$$(GeV^2)$ ~~& $\lambda$$(GeV)$ ~~& $\chi^{\pi}a$$(GeV^2)$ ~~&
$\Delta$$(GeV^4)$ ~~& $\lambda$$(GeV)$ ~~& $\chi^{\pi}a$$(GeV^2)$\\ \hline    
 0.200& 1.65& 0.096& $10^{-1}$& 1.83& 0.112   \\ 
 0.020& 1.55& 0.066& $10^{-4}$& 1.02& 0.069  \\ 
 0.002& 1.45& 0.058& $10^{-7}$& 0.83& 0.081  \\ \hline\hline
\end{tabular}
\end{center}
\end{table}

From Table.III, one can see that the pion vacuum susceptibilities are not sensitive to the model gluon propagators and the small value in comparison to Ref.[7] is robust within this model approach. Here we want to stress that Eq.(24) is convergent even if we take into account the asymptotic behavior of the gluon propagator. This is quite different from the calculation of chiral quark condensate(22). Taking into account the asymptotic behavior of the gluon propagator in Schwinger-Dyson equation, for $q^2\to\infty$, we have the proper asymptotic behavior for quark propagator
\begin{equation}
G(q^2)-G^{pert}(q^2)\rightarrow\frac{1}{(q^2)^2[lnq^2]^{1-\gamma_{m}}},
\end{equation}
which, numerically, will appear very close to zero(see Fig.1 and Fig.2) because it is powerlaw suppressed. However, for the chiral quark condensate in Eq.(22), we have
\begin{equation}
\langle\tilde{0}|:\bar{q}q:|\tilde{0}\rangle\propto\int^{\mu}_{0}sds\frac{1}{s^2[lns]^{1-\gamma_{m}}}\propto\left\{ln\mu\right\}^{\gamma_{m}},
\end{equation}
which is logarithmic divergent($\gamma_{m}$ is the mass anomalous dimension). From Eq.(27), it is easy to check that Eq.(24) is free of the UV divergence. This means that using the forms of Eq.(25) and (26) do not cause any problems in the calculation of pion vacuum susceptibility.

To summarize, in the present letter, a subtraction mechanism
which is consistent with the definition of vacuum condensate(the
vacuum expectation value of the normal production of quark
operator) in QCD sum rule, is adopted to calculate pion vacuum
susceptibility(employing Wick theorem and considering in addition to
the usual Nambu-Goldstone phase a perturbative phase of GCM).
Within this approach the pion vacuum susceptibility is free of UV
divergence. This is reasonable and what is to be expected. The results of pion susceptibility we obtained are much smaller than the previous estimation, which can be used as a reference for further study. Finally, we want to stress that GCM is not renormalizable. Therefore, the scale at which a condensate is defined in our approach is a typical hadronic scale, which is implicitly determined by the model gluon propagator $g^2D(q^2)$ and the solution of the rainbow DS equation(6,7). This situation is very similar to the determination of vacuum condensates in the instanton liquid model where the scale is set by the inverse instanton size[28,29].

\vspace*{0.2 cm}
\noindent{\large \bf Acknowledgments}

This work was supported in part by the National Natural Science Foundation
of China under Grant Nos 19975062, 10175033 and 10135030.

\vspace*{0.2 cm}
\noindent{\large \bf References}
\begin{description}
\item{[1]} M. Shifman, A. Vainshtein and V. Zakharov, Nucl. Phys.
{\bf B147}, 385 (1979).
\item{[2]}  L. Reinders, H. Rubinstein and S. Yazaki, Phys. Rep.
{\bf 127}, 1 (1985); S. Narison, QCD Spectral Sum Rules (World Scientific, Singapore
, 1989), and references therein.
\item{[3]} B. L. Ioffe and A. V. Smilga, Nucl. Phys. {\bf B232}, 109 (1984).
\item{[4]} I. I. Balitsky and A. V. Yung, Phys. Lett. {\bf B129}, 328 (1983).
\item{[5]} S. V. Mikhailov and A. V. Radyushkin, JETP Lett. {\bf 43},
712 (1986); Phys. Rev. {\bf D45}, 1754 (1992); A. P. Bakulev and A. V. Radyushkin, Phys.  Lett. {\bf B271}, 223 (1991).
\item{[6]} E. M. Henley, W. -Y. Hwang and L. S. Kisslinger, Phys. Lett. {\bf B367}, 21 (1996).
\item{[7]} M. B. Johnson and L. S. Kisslinger, Phys. Rev. {\bf D57},
2847 (1998).
\item{[8]} R. T. Cahill and C. D. Roberts, Phys. Rev. {\bf D32}, 2419 (1985); P. C. Tandy, Prog. Part. Nucl. Phys. 39, 117 (1997); R. T. Cahill and S. M. Gunner, Fiz. {\bf B7}, 17 (1998), and references therein.
\item{[9]} C. D. Roberts and A. G. Williams, Prog. Part. Nucl. Phys. {\bf 33}, 477 (1994), and references therein.
\item{[10]} M. R. Frank and T. Meissner, Phys. Rev {\bf C57}, 345 (1998).
\item{[11]} C. D. Roberts, R. T. Cahill, and J. Praschiflca, Ann. Phys. (N. Y.){\bf 188}, 20 (1988).
\item{[12]} P. Maris, C. D. Roberts, and P. C. Tandy, Phys. Lett {\bf B420}, 267 (1998).
\item{[13]} C. D. Roberts, R. T. Cahill, M. E. Sevior, and N. Ianella, Phys. Rev. {\bf 49}, 125 (1994). 
\item{[14]} M. R. Frank and T. Meissner, Phys. Rev {\bf C53}, 2410 (1996).
\item{[15]} R. T. Cahill and S. Gunner, Phys. Lett. {\bf B359}, 281 (1995); Mod. Phys. Lett. {\bf A10}, 3051 (1995).
\item{[16]} P. Maris and C. D. Roberts, Phys. Rev {\bf C56}, 3369 (1997).
\item{[17]} Xiao-fu L\"{u}, Yu-xin Liu, Hong-shi Zong and En-guang Zhao, Phys. Rev. {\bf C58}, 1195 (1998); Hong-shi Zong, Xiao-fu L\"{u}, Jian-zhong Gu, Chao-hsi Chang and En-guang Zhao, Phys. Rev. {\bf C60}, 055208 (1999); Hong-shi Zong, Xiao-fu L\"{u}, Fan Wang, Chao-hsi Chang, and En-guang Zhao, Commun Theor.Phys.(Beijing, China) {\bf 34}, 563 (2000); Hong-shi Zong, Yu-xin Liu, Xiao-fu L\"{u}, Fan Wang, and En-guang Zhao, Commun Theor. Phys.(Beijing, China) {\bf 36}, 187 (2001); Hong-shi Zong, Xiang-song Chen, Fan Wang, Chao-hsi Chang and En-guang Zhao, Phys. Rev. {\bf C66}, 015201 (2002).
\item{[18]} C. Burden, C. D. Roberts, and M. Thomson, Phys. Lett. {\bf B371}, 163 (1996).
\item{[19]} C. Burden and D. Liu, Phys. Rev. {\bf D55}, 367 (1997); M. A. Ivanov, Yu. L. Kalinovskii, P. Maris, and C. D. Roberts, Phys. Lett. {\bf B416}, 29 (1998); Phys. Rev. {\bf C57}, 1991 (1998). 
\item{[20]} A. Bender, D. Blaschke, Y. Kalinovskii, and C. D. Roberts, Phys. Rev. Lett. {\bf 77}, 3724 (1996).
\item{[21]} M. R. Frank, P. C. Tandy, and G. Fai, Phys. Rev {\bf C43}, 2808 (1991); M. R. Frank and P. C. Tandy, Phys. Rev {\bf C46}, 338 (1992); C. W. Johnson, G. Fai, and M. R. Frank, Phys. Lett. {\bf B386}, 75 (1996).
\item{[22]} R. T. Cahill, Nucl. Phys. {\bf A543}, 63 (1992).
\item{[23]} R. Alkofer and L. von Smekal, Phys. Rept. {\bf 353}, 281(2001); C. S. Fischer
and R. Alkofer, hep-ph/0301094, and references therein.
\item{[24]} P. Maris and C. D. Roberts, nucl-th/0301049, and references therein.
\item{[25]} T. Meissner, Phys. Lett. {\bf B405}, 8 (1997).
\item{[26]} Hong-shi Zong, Jia-lun Ping, Hong-ting Yang, Xiao-fu L\"{u} and Fan Wang, Phys. Rev. {\bf D67}, 074004 (2003).
\item{[27]} Hong-ting Yang, Hong-shi Zong, Jia-lun Ping, and Fan Wang, Phys. Lett. {\bf B557}, 33 (2003)
\item{[28]} M. V. Polyakov and C. Weiss, Phys. Lett. {\bf B387}, 841 (1996).
\item{[29]} A. E. Dorokhov, S. V. Esaibegian, and S. V. Mikhailov, Phys. Rev. {\bf D56}, 4062 (1997).
\end{description}

\end{document}